\documentclass[twocolumn]{article}
\usepackage{arxiv}
\usepackage{cite}
\usepackage{amsmath,amssymb,amsfonts}
\usepackage{algorithmic}
\usepackage{graphicx}
\usepackage{textcomp}

\usepackage{color,soul}
\usepackage[english]{babel}
\usepackage{hyperref}
\usepackage[utf8]{inputenc}
\usepackage[T1]{fontenc}
\usepackage{booktabs}
\usepackage{enumitem}
\definecolor{orange}{rgb}{0.85,0.15,0.15}
\definecolor{green}{rgb}{0.15,0.5,0.15}

\newcommand{\f}[1]{\textbf{#1}}
\newcommand{\s}[1]{\normalfont{#1}}

\usepackage[caption=false,font=footnotesize]{subfig}
\usepackage[labelfont=bf]{caption}
\usepackage[export]{adjustbox}
\usepackage{blindtext}
\usepackage{dblfloatfix}
\usepackage{makecell}

\providecommand{\keywords}[1]
{
  \small	
  \textbf{\textit{Keywords---}} #1
}

\begin{document}
\title{The Little W-Net That Could: \\State-of-the-Art Retinal Vessel Segmentation \\with Minimalistic Models}

\author{Adrian Galdran$^{1,*}$, André Anjos$^{2}$, José Dolz$^{3}$, Hadi Chakor$^{4}$, Hervé Lombaert$^{3}$, Ismail Ben Ayed$^{3,5}$  \\[3mm]
        $^{1}$University of Bournemouth, UK  \hspace{0.5cm}  $^{2}$IDIAP Research Institute, Switzerland \\[1mm]
         $^{3}$École de Technolgie Superieure de Montréal, Canada\hspace{0.5cm}  
         $^{4}$Diagnos Inc., Canada   \\[1mm] 
         $^{5}$CRCHUM University of Montreal Hospital Centre \\[1mm]
        $^{*}$ Corresponding author: agaldran@bournemouth.ac.uk
}

\twocolumn[
\begin{@twocolumnfalse}
\maketitle
\begin{abstract}
The segmentation of the retinal vasculature from eye fundus images represents one of the most fundamental tasks in retinal image analysis.
Over recent years, increasingly complex approaches based on sophisticated Convolutional Neural Network architectures have been slowly pushing performance on well-established benchmark datasets. 
In this paper, we take a step back and analyze the real need of such complexity. 
Specifically, we demonstrate that a minimalistic version of a standard U-Net with several orders of magnitude less parameters, carefully trained and rigorously evaluated, closely approximates the performance of current best techniques. 
In addition, we propose a simple extension, dubbed W-Net, which reaches outstanding performance on several popular datasets, still using orders of magnitude less learnable weights than any previously published  approach.
Furthermore, we provide the most comprehensive cross-dataset performance analysis to date, involving up to 10 different databases. 
Our analysis demonstrates that the retinal vessel segmentation problem is far from solved when considering test images that differ substantially from the training data, and that this task represents an ideal scenario for the exploration of domain adaptation techniques. 
In this context, we experiment with a simple self-labeling strategy that allows us to moderately enhance cross-dataset performance, indicating that there is still much room for improvement in this area. 
Finally, we also test our approach on the Artery/Vein segmentation problem, where we again achieve results well-aligned with the state-of-the-art, at a fraction of the model complexity in recent literature.
All the code to reproduce the results in this paper is released.
\end{abstract}
\vspace{0.15cm}
\vspace{0.5cm}
\end{@twocolumnfalse}
]

\begin{figure*}[pt]
\centering
\subfloat[]{\includegraphics[width = 0.19\textwidth,valign=c]{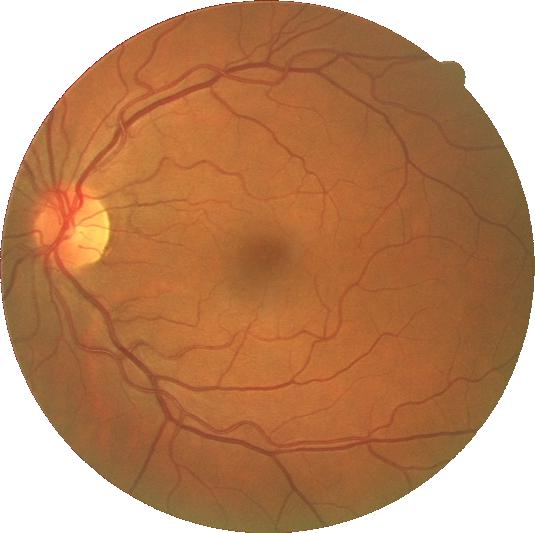}
\label{fig_data_1}}
\hfil
\subfloat[]{\includegraphics[width = 0.19\textwidth,valign=c]{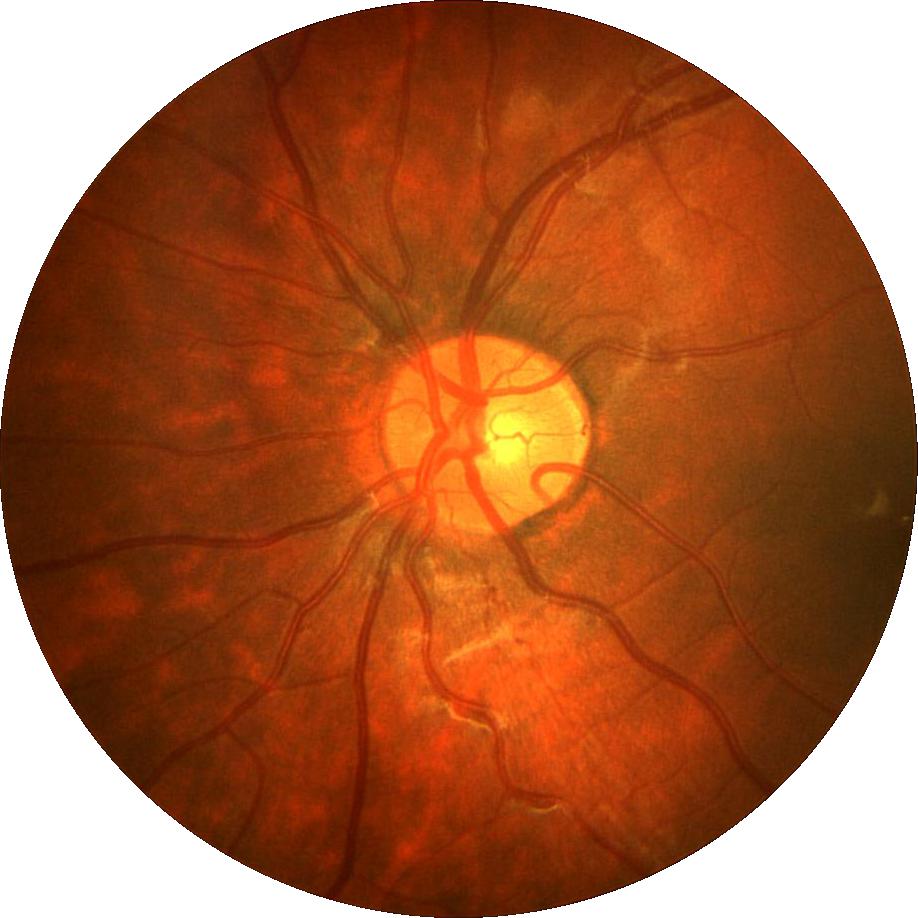}
\label{fig_data_2}}
\hfil
\subfloat[]{\includegraphics[width = 0.19\textwidth,valign=c]{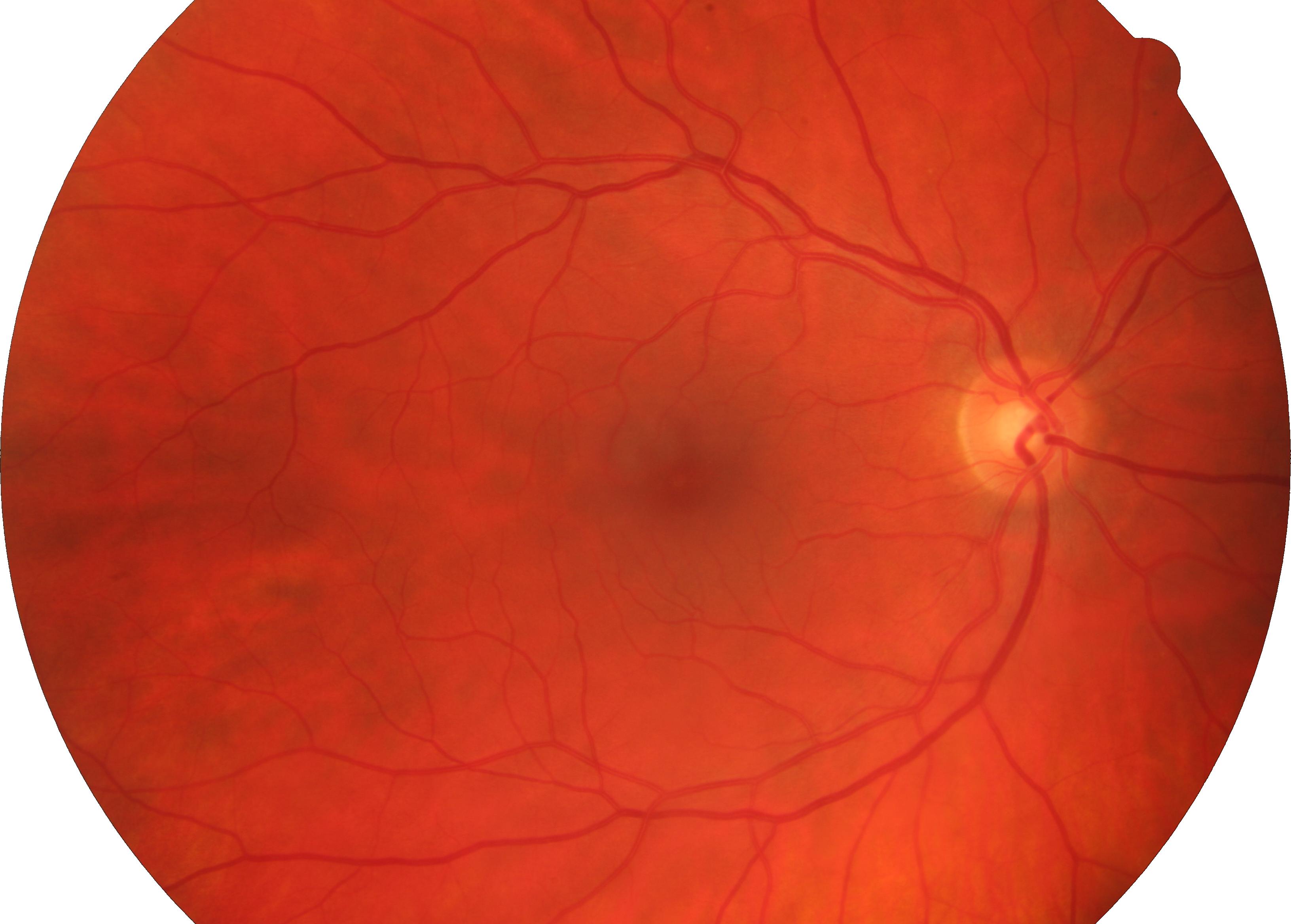}
\label{fig_data_2}}
\hfil
\subfloat[]{\includegraphics[width = 0.19\textwidth,valign=c]{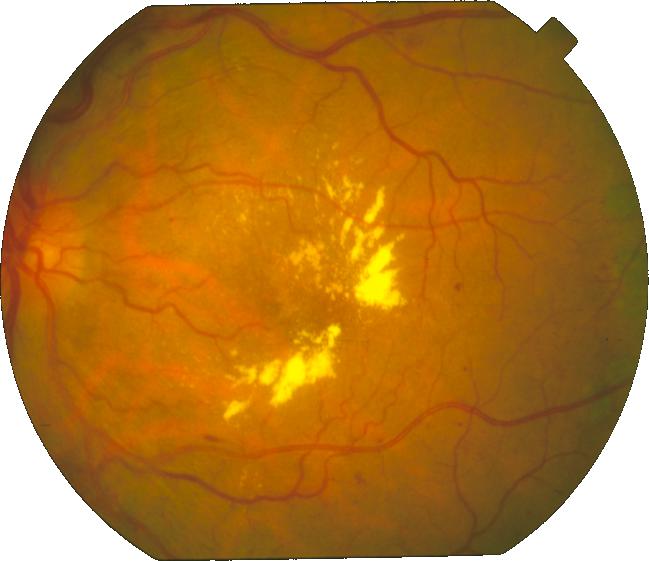}
\label{fig_data_2}}
\hfil
\subfloat[]{\includegraphics[width = 0.19\textwidth,valign=c]{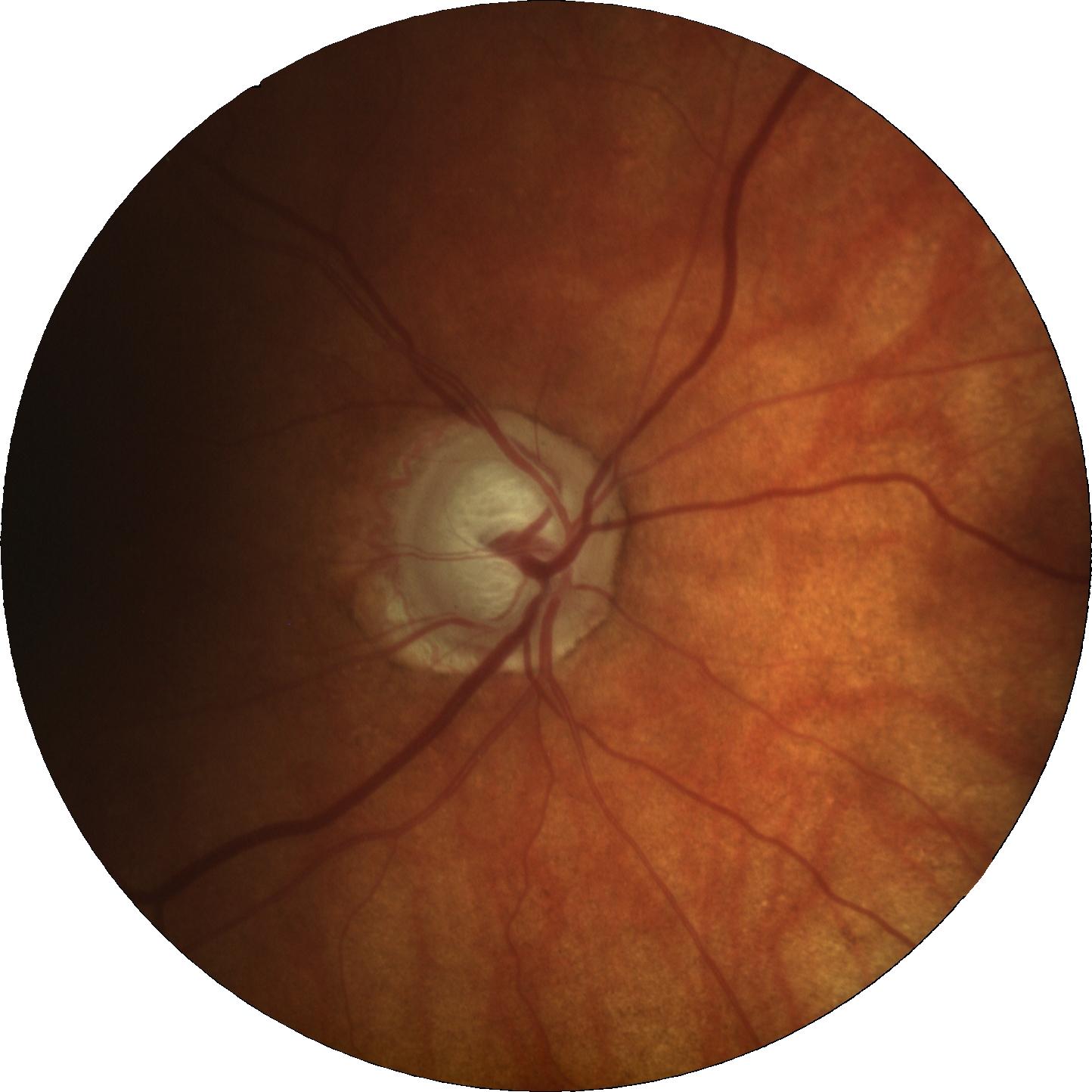}
\label{fig_data_2}}

\subfloat[]{\includegraphics[width = 0.19\textwidth,valign=c]{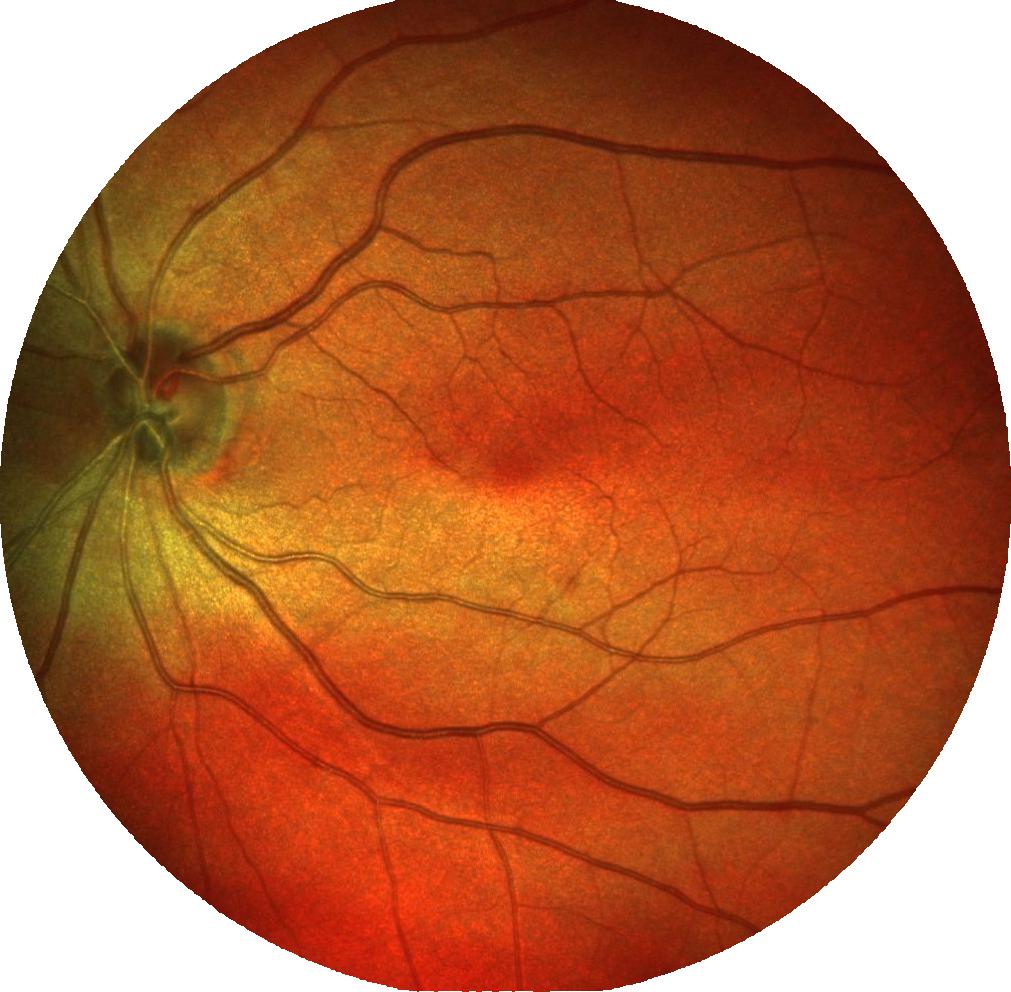}
\label{fig_data_1}}
\hfil
\subfloat[]{\includegraphics[width = 0.19\textwidth,valign=c]{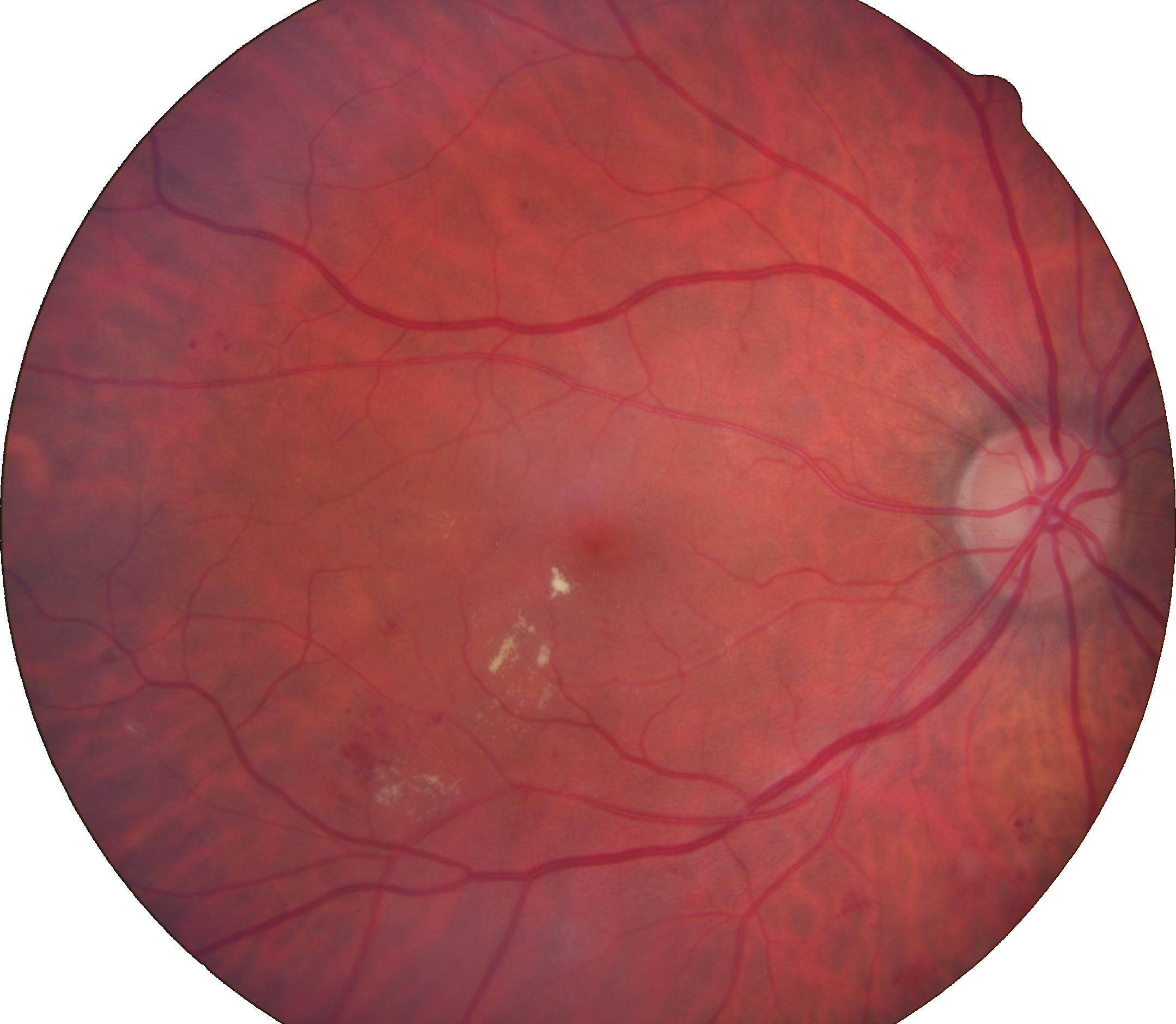}
\label{fig_data_2}}
\hfil
\subfloat[]{\includegraphics[width = 0.19\textwidth,valign=c]{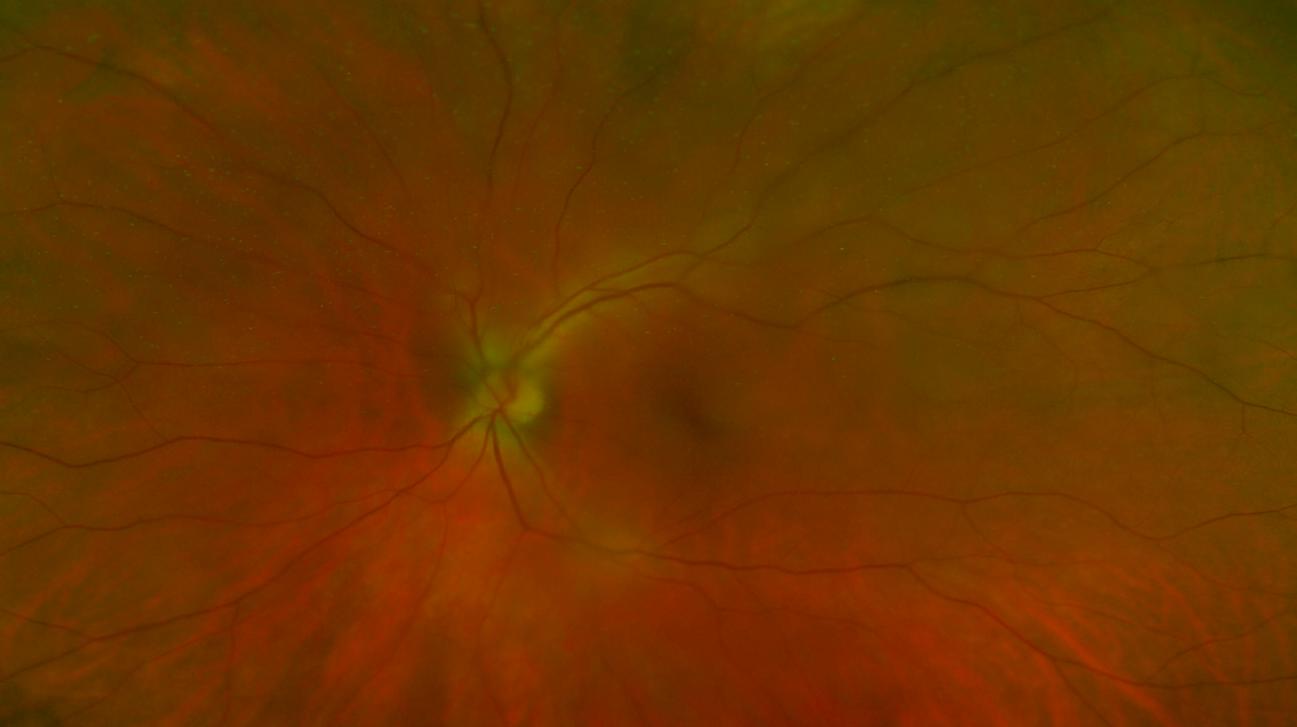}
\label{fig_data_2}}
\hfil
\subfloat[]{\includegraphics[width = 0.19\textwidth,valign=c]{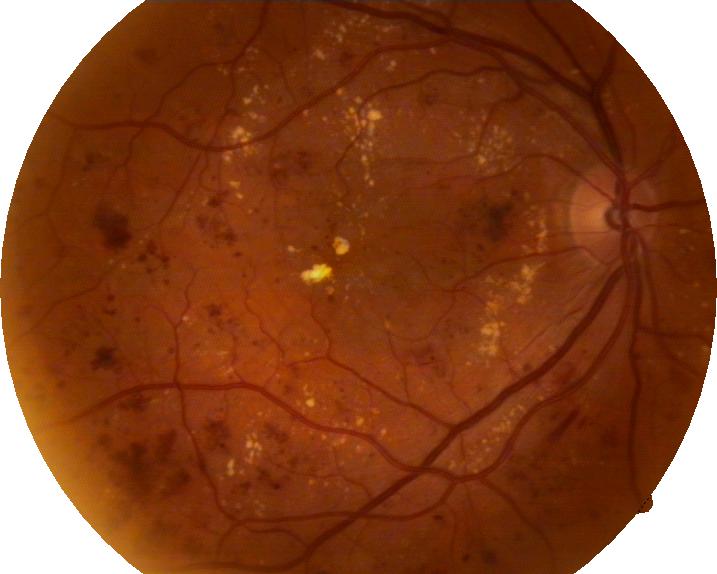}
\label{fig_data_2}}
\hfil
\subfloat[]{\includegraphics[width = 0.19\textwidth,valign=c]{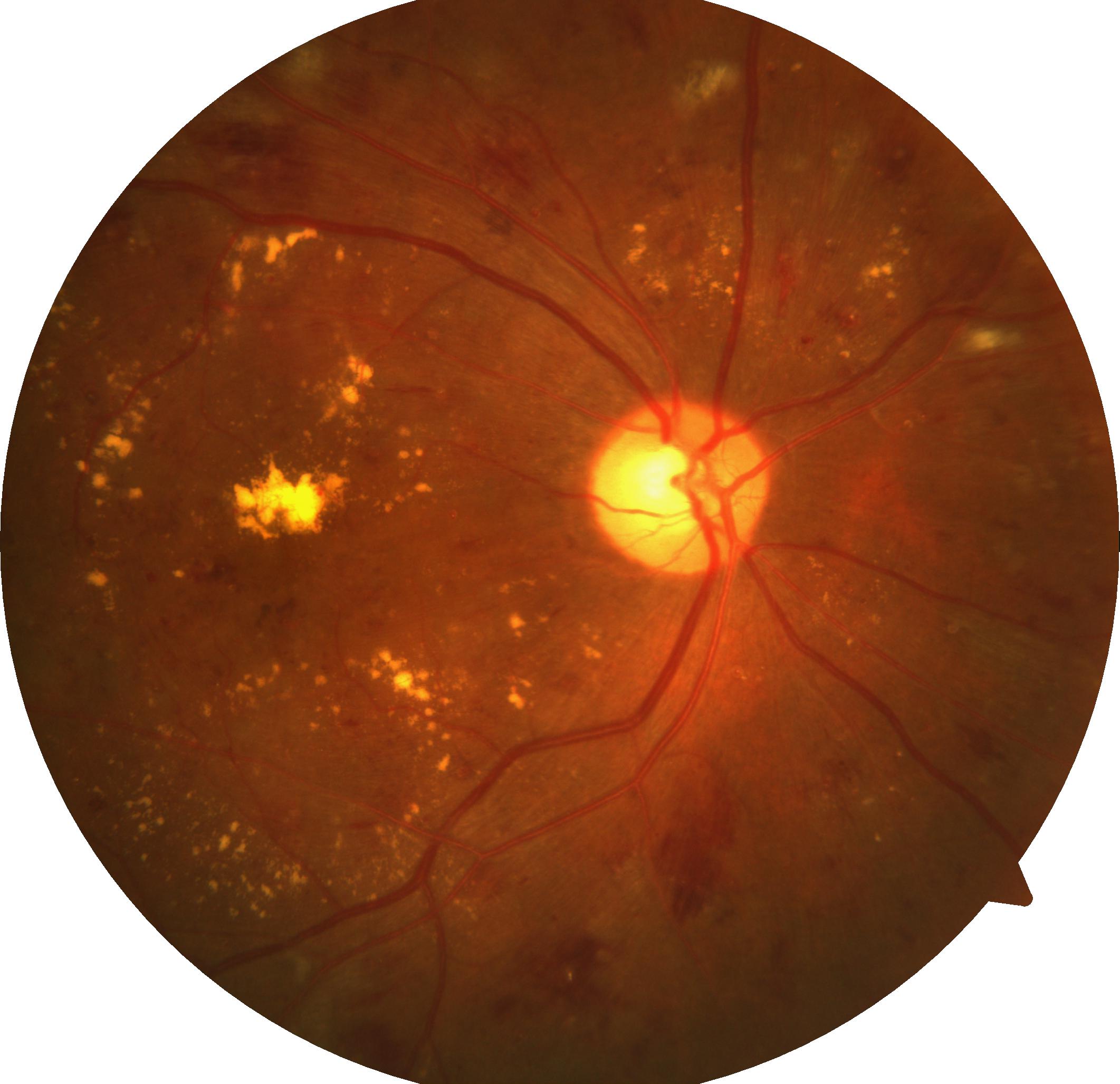}
\label{fig_data_2}}
\caption{This work provides the most comprehensive cross-dataset performance study on vessel segmentation to date. A representative image from each of the 10 databases considered in this paper: DRIVE \cite{staal_ridge-based_2004}, CHASE-DB 1 \cite{fraz_ensemble_2012}, HRF \cite{wang_robust_2013}, STARE \cite{hoover_locating_2000}, LES-AV \cite{orlando_towards_2018}, IOSTAR \cite{zhang_robust_2016}, DR HAGIS \cite{holm_dr_2017}, AV-WIDE \cite{estrada_retinal_2015}, DRIDB \cite{prentasic_diabetic_2013}, UoA-DR  \cite{chalakkal_comparative_2017}. A detailed description of each database is given in Table \ref{tab_results}.}
\label{fig_datasets}
\end{figure*}

\section{Introduction and Related Work}\label{sec:introduction}
Retinal vessel segmentation is one of the first and most important tasks in for the computational analysis of eye fundus images. 
It represents a stepping stone for more advanced applications like artery/vein ratio computation \cite{niemeijer_automated_2011}, blood flow analysis \cite{orlando_towards_2018}, image quality assessment \cite{welikala_automated_2016}, retinal image registration \cite{chen_retinal_2015}, or retinal image synthesis \cite{costa_end--end_2018}.

Initial approaches to retinal vessel segmentation were fully unsupervised and relied on conventional image processing operations like mathematical morphology \cite{zana_segmentation_2001,mendonca_segmentation_2006} or adapted edge detection operations \cite{frangi_multiscale_1998}. 
The idea behind these methods is to apply some kind of transformation to a retinal image so that vessel intensities are emphasized, and then threshold the result to achieve a segmentation.
Although research on advanced filtering techniques for retinal vessel segmentation has continued over more recent years \cite{azzopardi_trainable_2015, zhang_robust_2016}, these kind of techniques consistently reach lower performance on established benchmarks, likely due to their inability to handle images with pathological structures and generalizing to images with different appearances and resolutions.

In contrast, early learning-based approaches quickly showed more promising results and better performance than their image processing counterparts \cite{staal_ridge-based_2004,soares_retinal_2006,marin_new_2011,becker_supervised_2013,orlando_discriminatively_2017}. The common strategy of these techniques consisted on the extraction of specifically designed local descriptors that were later passed to a relatively simple classifier, and the focus became to derive the most discriminative visual features for the task at hand. 

This predominance of machine learning techniques was reinforced with the emergence of deep neural networks. 
After initial realization that Convolutional Neural Networks (CNNs) could outperform previous methods, bypassing any manual feature engineering and directly learning from raw data \cite{liskowski_segmenting_2016,maninis_deep_2016}, a constant stream of publications has kept appearing on this topic, up to the point that almost any new competitive vessel segmentation technique is based now on this approach.

Standard CNN approaches to retinal vessel segmentation are based on the sequential application of a stack of convolutional layers that subsequently downsample and upsample input images to reach a probabilistic prediction of vessel locations. 
The weights of the network are then iteratively updated to improve those predictions by means of the minimization of a given miss-classification loss, \textit{e.g.} cross-entropy. 
Either processing small image patches \cite{liskowski_segmenting_2016} or the entire image \cite{maninis_deep_2016}, these approaches can succeed in segmenting the retinal vasculature with few annotated training data.

Extensions to the above paradigm tend to involve complex operations, like specifically designed network layers. 
Fu \textit{et al.} introduced a Conditional Random Field recurrent layer to model more global relationships between pixels \cite{fu_deepvessel_2016}, and Shi \textit{et al.} combined convolutional and graph-convolutional layers to better capture global vessel connectivity \cite{shin_deep_2019}. Guo \textit{et al.} introduced dense dilated layers that adjust the dilation rate based on vessel thickness \cite{guo_bscn_2020}, and Fan \textit{et al.} proposed a multi-frequency convolutional layer (OctConv) in \cite{fan_accurate_2019}.
Other custom convolutional blocks and layers based on domain knowledge have been explored in several recent works \cite{wang_ctf-net_2020,cherukuri_deep_2020}. 

Non-standard losses have also been proposed in recent years. Yan \textit{et al.} \cite{yan_joint_2018} trained a U-Net architecture \cite{ronneberger_u-net_2015} by minimizing a joint-loss that receives output predictions from two separate network branches, one with a pixel-level and one with a segment-wise loss. The same authors introduced a similar segment-level approach in \cite{yan_three-stage_2019}, whereas Mou \textit{et al.} employed a multi-scale Dice loss \cite{mou_dense_2020}, Zhao \textit{et al.} proposed a combination of global pixel-level loss and local matting loss \cite{zhao_improving_2020}, and Zhang and Chung introduced in \cite{zhang_deep_2018} a deeply supervised approach in which various loss values extracted at different stages of a CNN are combined and backpropagated, with artificial labels in vessel borders turning the problem into a muti-class segmentation task.
Generative Adversarial Networks have also been proposed for retinal vessel segmentation \cite{lahiri_generative_2017,son_towards_2019,zhao_supervised_2019,park_m-gan_2020}, although without achieving widespread popularity due to the difficulty in training these techniques. 

It is also worth reviewing efficient approaches to retinal vessel segmentation, as we plan to introduce in this paper high-performance lightweight models. 
These methods typically appear in works focused on retinal vessel segmentations for embedded/mobile devices
In this context, conventional unsupervised approaches are still predominant.
Arguello \textit{et al.} employ image filtering coupled with contour tracing \cite{arguello_gpu-based_2018},  Bibiloni \textit{et al.} apply simple hysteresis thresholding \cite{bibiloni_real-time_2019}, whereas Xu \textit{et al.} adapt Gabor filters and morphological operations for vessel segmentation in smartphone devices \cite{xu_smartphone-based_2016}. 
Only recently, Laibacher \textit{et al.} have explored efficient CNN architectures specifically designed for vessel segmentation on eye fundus images \cite{laibacher_m2u-net_2019}. Their proposed M2U-Net architecture leverages an ImageNet-pretrained MobileNet model \cite{sandler_mobilenetv2_2018} and achieves results slightly inferior to the state-of-the-art.
\subsection{Goals and Contributions}
The goal of this paper is to show that 1) as recently shown in other computer vision problems \cite{chen_closer_2019,musgrave_metric_2020}, there is no need of designing complex CNN architectures to outperform \textit{most current techniques} on the task of retinal vessel segmentation, and 2) when a state-of-the-art model is trained on a particular dataset and tested on images from different data sources, it can result in poor performance. 
On our way to establish these two facts, we make several contributions:
\begin{enumerate}
\item We introduce a simple extension of the standard U-Net architecture, named W-Net, which allows us to achieve outstanding performance on well-established datasets.
\item We establish a rigorous evaluation protocol, aiming to correct previous pitfalls in the area.
\item We test our approach in a large collection of retinal datasets, consisting of 10 different databases showing a wide range of characteristics, as illustrated in Fig. \ref{fig_datasets}.
\item Our cross-dataset experiments reveal that domain shift can induce performance degradation in this problem. We propose a simple strategy to address this challenge, which is shown to recover part of the lost performance.
\item Finally, we also apply our technique to the related problem of Artery/Vein segmentation from retinal fundus images, matching the performance of previous approaches with models that contain much fewer parameters.
\end{enumerate}

We believe that our results open the door to a more systematic study of new domain adaptation techniques in the area of retinal image analysis: since training one of our models to reach superior performance takes approximately 20 minutes in a single consumer GPU, our work can serve as a first step for quick design and experimentation with improved approaches that can eventually bridge the generalization gap across different data sources revealed by our experiments. 
To favor research in this direction, we release the code and data to reproduce our results at \url{github.com/agaldran/lwnet}.

\begin{figure*}[t]
\centering
\includegraphics[width = 0.99\textwidth]{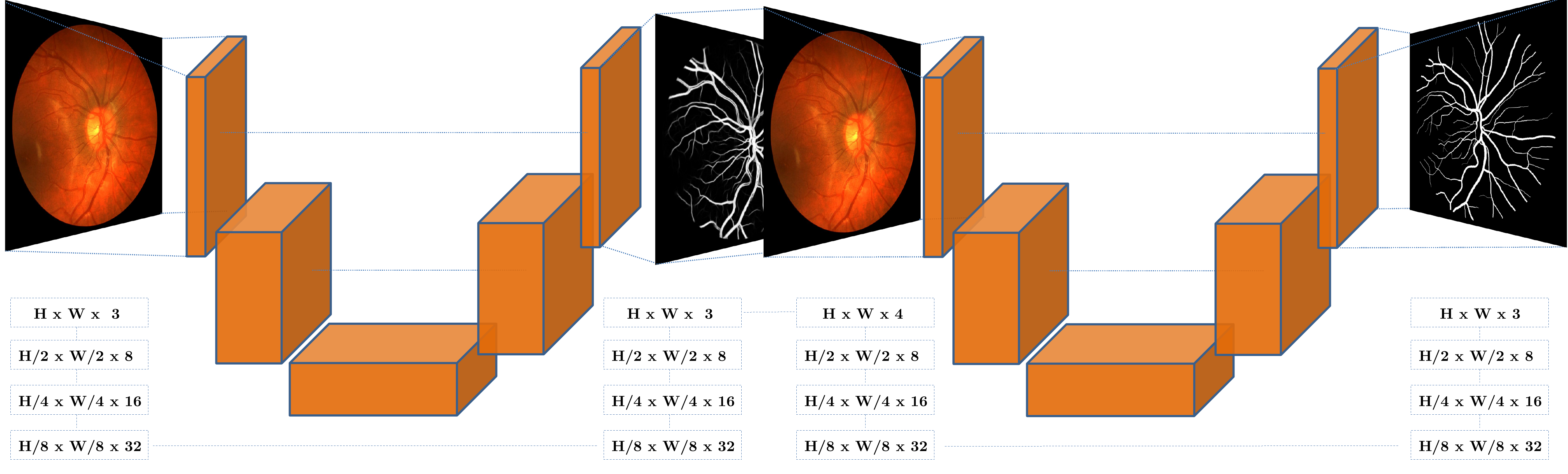}
\caption{Representation of the WNet architecture. The left-hand-side part of the architecture corresponds to a standard minimal U-Net $\phi_{3,8}$ with $\sim$34K parameters, and it achieves performance on-par with the state-of-the-art. The full W-Net, defined by eq. (\ref{wnet_def}), is composed of two consecutive U-Nets; it outperforms all previous approaches with just around 70k parameters: 1-3 orders of magnitude less than previously proposed CNNs.}\label{fig_wnet}
\end{figure*}

\section{Methodology}

\subsection{Baseline U-Net: structure and complexity}
One of the main goals of this work is to explore the lower limits in model complexity for the task of retinal vessel segmentation. 
Accordingly, we consider one of the simplest and most popular architectures in the field of medical image segmentation, namely the U-Net \cite{ronneberger_u-net_2015}. 
A standard U-Net is a convolutional autoencoder built of a downsampling CNN that progressively applies a set of filters to the input data while reducing its spatial resolution, followed by an upsampling path that recovers the original size.
U-Nets typically contain skip connections that link activation volumes from the downsampling path to the upsampling path via concatenation or addition in order to recover higher resolution information and facilitate gradient flow during training.

Let us parametrize a U-Net architecture $\phi$ by the number of times the resolution is downscaled/upscaled $k$, and the number of filters applied in each of these depth levels, $f_k$. 
To simplify our analysis, we will only consider filters of size $3\times3$, and we double the amount of filters each time we increase $k$ - this is a common pattern in U-Net designs.
Therefore, in this work a U-Net is fully specified by a pair of numbers ($k$, $f_0$), and we denote it by $\phi_{k,f_0}$.
In addition, we assume that Batch-Norm layers are inserted after each convolutional operation and that extra skip connections are added within each block. 
An example of such design pattern is shown in the left hand side of Fig. \ref{fig_wnet}.
In the remaining of this work, we will consider. the $\phi_{3,8}$ architecture, which contains approximately $34,000$ parameters. It is important to stress that this represents 1-3 orders of magnitude less than previously proposed CNNs for the task or retinal vessel segmentation.

\subsection{The W-Net architecture}
We also introduce a modification of the U-Net architecture, that we refer to as W-Net. 
The idea behind a W-Net, denoted by $\Phi$, is straightforward: for an input image $x$, the result of forward-passing it through a standard U-Net $\phi^1(x)$ is concatenated to $x$, and passed again through a second U-Net, which would be represented as:
\begin{equation}\label{wnet_def}
\Phi(x) = \phi^2(x, \phi^1(x)).
\end{equation}
In practice, $\phi^1$ generates a first prediction of vessels localization that can then be used by $\phi^2$ as a sort of attention map to focus more on interesting areas of the image, as shown in Fig. \ref{fig_wnet}. 
Of course a W-Net $\Phi$ contains twice the amount of learnable parameters as a standard U-Net. 
However, since the base U-Nets $\phi_{3,8}^1, \phi^2_{3,8}$ involved in its definition contain only $34,000$ each, the W-Net considered in this paper will have around $68,000$ weights, which is still one order of magnitude below the simplest architecture proposed to date for vessel segmentation, and three orders of magnitude smaller than state-of-the-art architectures.

\begin{figure*}[b]
\centering
\includegraphics[width = 0.95\textwidth]{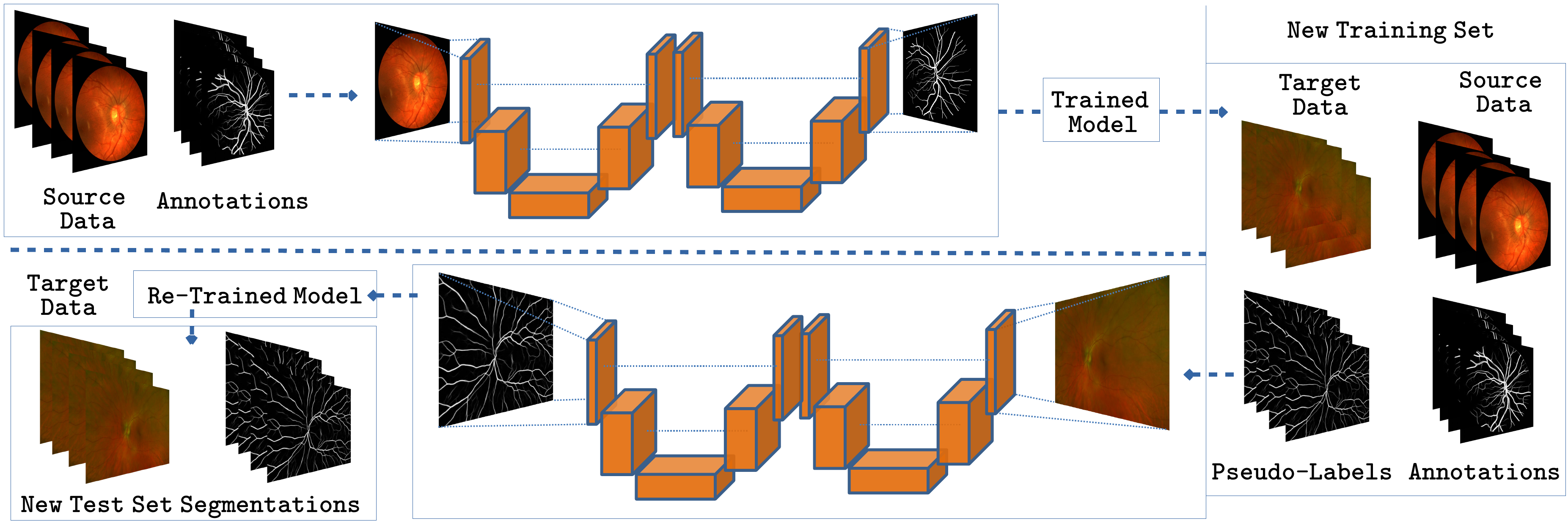}
\caption{Domain Adaptation strategy employed in this work: A model trained on source data is used to generate pseudo-labels on a target dataset. The original source data and the target data with the pseudo-labels are used to fine-tune that model and produce better predictions.}\label{da_fig}
\end{figure*}

\subsection{Training Protocol}
In all the experiments reported in this paper, the training strategy remains the same. 
Specifically, we minimize a standard cross-entropy loss between the predictions of the model on an image $x$ and the actual labels $y$. 
It is worth mentioning that in the W-Net case, an auxiliary loss is computed for the output of the first network and linearly combined with the loss computed for the second network:
\begin{equation}
\mathcal{L}(\Phi(x), y) = \mathcal{L}(\phi^1(x), y) + \mathcal{L}(\phi^2(x), y)
\end{equation}
The loss is backpropagated and minimized by means of the Adam optimization technique. 
The learning rate is initially set to $\lambda=10^{-2}$, and cyclically annealed following a cosine law until it reaches $\lambda=10^{-8}$.
Each cycle runs for $50$ epochs, and we adjust the amount of cycles (based on the size of each training set) so that we reach $4000$ iterations in every experiment. 

Images are all resized to a common resolution and processed with standard data augmentation techniques, and the batch size is set to $4$ in all experiments.
During training, at the end of each cycle the Area Under the ROC curve is computed on a separate validation set, and the best performing model is kept. 
Test-Time-Augmentations (horizontal and vertical image flips) are applied during inference in all our experiments.

\subsection{A simple Baseline for Domain Adaptation}\label{da}
One of the main goals in this paper is to show that, even if simple approaches can outperform much more complex current techniques, the problem of retinal vessel segmentation is not as trivial as we may extrapolate from this. 
The reason is that models trained on a given dataset do not reach the same level of performance when tested on retinal images sampled from markedly different distributions, as we quantitatively show in section \ref{da_eval}. A relevant drop of performance appears when a model trained on a given source dataset $\mathcal{S}$ is used to generate segmentations on a substantially different target dataset $\mathcal{T}$. 

Attempting to close such performance gap is a task falling within the area of Domain Adaptation, which has been subject of intensive research in the computer vision community for the last years \cite{kouw_review_2019}. 
Here we explore a simple solution to address this challenge in the context of retinal vessel segmentation. 
Namely, given a model $U_\mathcal{S}$ trained on $\mathcal{S}$ we proceed by first generating probabilistic  segmentations for each image $x\in\mathcal{T}$. We then merge the source dataset labels $y_\mathcal{S}$ with the target dataset segmentations $\{U_\mathcal{S}(x) \ | \ x \in \mathcal{T}\}$, which we treat as pseudo-labels.
Lastly, we fine-tune $U_\mathcal{S}$ in this new dataset, starting from the weights of the model trained on $\mathcal{S}$, with a learning rate reduced by a factor of $100$, for $10$ extra epochs. 
During training, we monitor the AUC computed in the training set (including both source labels and target pseudo-labels) as a criterion for selecting the best model. 
It is worth stressing that pseudo-labels $U_\mathcal{S}(x)$ remain with probabilistic values in $[0,1]$, rather than binary, with the goal of informing the model about the uncertainty present on them. The rationale behind this is to force the new model to learn from segmentations in $\mathcal{S}$ with confident annotations, while at the same time exposing it to images from $\mathcal{T}$ before testing on them. A graphical overview of this strategy is shown in Fig. \ref{da_fig}.

\begin{table*}[t]  %
	\renewcommand{\arraystretch}{1.3}	
	\centering
\setlength\tabcolsep{4pt}	
\begin{tabular}{lc  cc cc}
				                          & \bfseries \textbf{Year}  & \textbf{\# ims.}  &  \textbf{Resolution}       &  \textbf{FOV}   &  \textbf{Challenges \& Comments}    \\
\midrule
STARE \cite{hoover_locating_2000}         & 2000 				 &  20&  605$\times$700     & 35$^{\circ}$   &   
\makecell{Poor quality: scanned and digitized photographs\\Healthy and pathological images (10/10)} \\
\midrule
DRIVE \cite{staal_ridge-based_2004}       & 2004 				&  40 &  565$\times$584     & 45$^{\circ}$         &   
\makecell{Consistent good quality and contrast, low resolution  \\Mostly healthy patients, some with mild DR (33/40)} \\
\midrule
CHASE-DB 1 \cite{fraz_ensemble_2012}      & 2012 				&  28 &  999$\times$960         & 30$^{\circ}$          &   
\makecell{OD-centered images from 10-year old children\\Uneven background illumination and poor contrast} \\
\midrule
HRF \cite{wang_robust_2013}               & 2013 				&  45&  3504$\times$2336          &  60$^{\circ}$     &   
\makecell{High visual quality, images taken with mydriatic dilation \\Healthy, diabetic, and glaucomatous patients (15/15/15)} \\
\midrule
DRiDB \cite{prentasic_diabetic_2013}      & 2013 				&  50 &  720$\times$576         &  45$^{\circ}$        &   
\makecell{Highly varying quality, illumination, and image noise \\Mostly diabetic patients of varying grades (36/50)} \\
\midrule
AV-WIDE \cite{estrada_retinal_2015}       & 2015 				&  30 &   \makecell{2816$\times$1880\\1500$\times$900}  & 200$^{\circ}$ &   
\makecell{Uneven illumination, varying resolution due to cropping \\Healthy and age-related macular degeneration patients.} \\
\midrule
IOSTAR \cite{zhang_robust_2016}           & 2016 				&  30&   1024$\times$1024   & 45$^{\circ}$   &   
\makecell{Scanning Laser Ophthalmoscope images \\ Macula-centered, high contrast and visual quality} \\
\midrule
DR HAGIS \cite{holm_dr_2017}              & 2017 				 &  40&  \makecell{2816$\times$1880\\4752$\times$3168}   & 45$^{\circ}$         &   
\makecell{Multi-center, multi-device macula-centered images\\All diabetic patients with different co-morbities} \\
\midrule
UoA-DR \cite{chalakkal_comparative_2017}  & 2017 				&  200 &  2124$\times$2056     &  45$^{\circ}$     &   
\makecell{Both macula and OD-centered images \\Healthy, NP-DR and P-DR patients (56/114/30)} \\
\midrule
LES-AV \cite{orlando_towards_2018}        & 2018 				&  22 &  \makecell{1144$\times$1620\\1958$\times$2196}     & \makecell{30$^{\circ}$\\45$^{\circ}$}  &
\makecell{OD-centered images, highly varying illumination \\11 healthy and 11 glaucomatous patients} \\
\bottomrule
\end{tabular}
\caption{Description of each of the ten datasets considered in this paper in terms of image and population characteristics.}
\label{tab_datasets}
\end{table*}%

\subsection{Evaluation Protocol}\label{eval_protocol}
Unfortunately, a rigorous evaluation protocol for retinal vessel segmentation is missing in the literature due to several issues: differences in train/test splits in common benchmarks, or wrongly computed performance metrics. 
Below we outline what we understand as a correct evaluation procedure:
\begin{enumerate}[leftmargin=*]
\item  All performance metrics are computed at native image resolution and excluding pixels outside the retinal area, which are trivially predicted as having zero probability of being part of a vessel.
\item Whenever an official train/test split exists, we follow it. When there is none, we follow the least ``favorable'' split we could find in previous works, \textit{i.e.} the one assigning less images for training. 
We make this decision based on the low difficulty of the vessel segmentation task; this is in contrast with other works that employ leave-one-out cross-validation, which can use up to $95\%$ of the data for training \cite{oliveira_retinal_2018,yan_joint_2018}.
\item We first accumulate all probabilities and labels across the training set, then perform AUC analysis and derive an optimal threshold (maximizing the Dice score) to binarize predictions. We then apply the same procedure on the test set, now using the pre-computed threshold to binarize test segmentations. This stands opposed to computing metrics per-image and reporting the mean performance \cite{xu_improved_2017}, or using a different threshold on each test image for binarizing probabilistic predictions \cite{zhuo_size-invariant_2020}. 
\item Cross-dataset experiments are reported in a variety of different datasets. No pre-processing or hyper-parameters are re-adjusted when changing datasets, since this heavily undermines the utility of a method. This is a typical shortcoming of unsupervised approaches, which tend to modify certain parameters to account for different vessel calibers \cite{zhang_robust_2016}. Also, the threshold to binarize predictions on different datasets is the one derived from the original training set, without using test data to readjust it. 
\item We do not report accuracy, since this is a highly imbalanced problem; the Dice score is a more suitable figure of merit. We also report Matthews Correlation Coefficient (MCC), as it is better suited for imbalanced problems \cite{chicco_advantages_2020}. Sensitivity and specificity computed at a particular cut-off value are avoided, as they are useless when comparing the performance of different models. 
\end{enumerate}

\section{Experimental Results}
In this section we provide a comprehensive performance analysis of the methodology introduced above.

\subsection{Datasets Description}
A key aspect of this work is our performance analysis of a wide range of data sources. 
For each of the considered models, we train them on three different datasets, namely DRIVE \cite{staal_ridge-based_2004}, CHASE-DB \cite{fraz_ensemble_2012} and HRF \cite{wang_robust_2013}. 
The train/validation/test splits for DRIVE are provided by the authors, but there is no official split in the other two cases. 
We decide to adopt the most restrictive splits we could find in the literature \cite{laibacher_m2u-net_2019}: only 8 of the 22 images in CHASE-DB, and 15 of the 45 images in HRF are used for training and validation.

After training, we test our models on the corresponding test sets. 
In section \ref{da_eval}, we also consider another seven different datasets for cross-datasets and domain adaptation evaluation. 
These include a variety of different image qualities, resolutions, pathologies, and even image modalities.
Further details of each of these databases are given in Table \ref{tab_datasets}. 

It is also worth mentioning that, for training, all images from DRIVE, CHASEDB, and HRF are downsampled to a $512\times512$, $512\times512$, and $1024\times1024$ resolution respectively, whereas evaluation is carried out at native resolution for all datasets. No pre-processing (nor post-processing) was applied.

\begin{table*}[h]  %
	\renewcommand{\arraystretch}{1}	
	\centering
\setlength\tabcolsep{3pt}	
\begin{tabular}{lccc  ccc cc ccc}
& &  & \multicolumn{3}{c}{\textbf{DRIVE}} & \multicolumn{3}{c}{\textbf{CHASE-DB}} & \multicolumn{3}{c}{\textbf{HRF}} \\
\cmidrule(lr){4-6} \cmidrule(lr){7-9} \cmidrule(lr){10-12} 
				& \bfseries \# Pub/Year&   \bfseries \# Params                                &  AUC  &  DICE &  MCC &  AUC &  DICE &  MCC   &   AUC   &  DICE &  MCC   \\
\midrule
Maninis \textit{et al.} \cite{maninis_deep_2016}            &\textbf{ECCV/2016}  &            &  ---  & 82.20 &  ---&  ---  &  ---  &  ---   &   ---   &  ---  &  --- \\
Zhang \textit{et al.} \cite{zhang_robust_2016}              &\textbf{TMI/2016}   &            & 96.36 &  ---  &  ---& 96.06 &  ---  &  ---   &  96.08  &   --- & 74.10  \\
Fu \textit{et al.} \cite{fu_deepvessel_2016}                &\textbf{MICCAI/2016}&            & 94.04 & 78.75 &  ---& 94.82 & 75.49 &  ---   &   ---   &   --- &  ---   \\
Liskowski \textit{et al.} \cite{liskowski_segmenting_2016}  &\textbf{TMI/2016}   & 48,000,000 & 97.90 &  ---  &  ---& 98.45 &  ---  &  ---   &   ---   &   --- &  ---   \\
Orlando \textit{et al.} \cite{orlando_discriminatively_2017}&\textbf{TBME/2017}  &            & 95.07 & 78.57 &75.56& 95.24 & 73.32 & 70.46  &  95.24  & 71.58 & 68.97 \\
Gu {et al.} \cite{gu_segment_2017}                          &\textbf{TMI/2017}	 &            &  ---  & 78.86 &75.89&  ---  & 72.02 & 69.08  &   ---   & 77.49 & 75.41  \\
Wu \textit{et al.} \cite{wu_multiscale_2018}                &\textbf{MICCAI/2018}&            & 98.07 &  ---  &  ---& 98.25 &  ---  &  ---   &   ---   &   --- &  ---   \\
Yan \textit{et al.} \cite{yan_joint_2018}                   &\textbf{TBME/2018}  &            & 97.52 & 81.83 &  ---& 97.81 &  ---  &  ---   &   ---   & 78.14 &  ---  \\
Wang \textit{et al.} \cite{wang_retinal_2019}               &\textbf{BSPC/2019}  &            &  ---  & 81.44 &78.95&  ---  & 78.63 & 76.55  &   ---   &   --- &  ---   \\
Wang \textit{et al.} \cite{wang_dual_2019}                  &\textbf{MICCAI/2019}&            & 97.72 & 82.70 &  ---& 98.12 & 80.37 &  ---   &   ---   &   --- &  ---   \\
Araujo \textit{et al.} \cite{araujo_deep_2019}              &\textbf{MICCAI/2019}&            & 97.90 &  ---  &  ---& 98.20 &  ---  &  ---   &   ---   &   --- &  ---   \\
Fu \textit{et al.} \cite{fu_divide-and-conquer_2019}        &\textbf{MICCAI/2019}&            & 97.19 & 80.48 &  ---&  ---  &  ---  &  ---   &   ---   &   --- &  ---   \\
Wang \textit{et al.} \cite{wang_blood_2019}                 &\textbf{PatRec/2019}&            &  ---  & 80.93 &78.51&  ---  & 78.09 & 75.91  &   ---   & 77.31 &  ---   \\
Wu \textit{et al.} \cite{gu_ce-net_2019}                    &\textbf{TMI/2019}   &            & 97.79 &  ---  &  ---&  ---  &  ---  &  ---   &   ---   &   --- &  ---   \\
Zhao \textit{et al.} \cite{zhao_supervised_2019}            & \textbf{TMI/2019}  &            &  ---  & 78.82 &  ---&  ---  &  ---  &  ---   &   ---   & 76.59 &  ---    \\
Laibacher \textit{et al.} \cite{laibacher_m2u-net_2019}      &\textbf{CVPR-W/2019}& 549,748   & 97.14 & 80.91 &  ---& 97.03 & 80.06 &  ---   &   ---   & 78.14 &  ---  \\
Shin \textit{et al.} \cite{shin_deep_2019}                  &\textbf{MedIA/2019} & 7,910,000  & 98.01 & 82.63 &  ---& 98.30 & 80.34 &  ---   &  \textbf{98.38} &  \textbf{81.51} &  --- \\
Zhao \textit{et al.}  \cite{zhao_improving_2020}            &\textbf{PatRec/2020}&            &  ---  & 82.29 &  ---&  ---  &  ---  &  ---   &   ---   & 77.31 &  ---   \\
Zhuo \textit{et al.} \cite{zhuo_size-invariant_2020}        &\textbf{CMPB/2020}  &            & 97.54 & 81.63 &  ---&  ---  &  ---  &  ---   &   ---   &   --- &  ---   \\
Mou \textit{et al.} \cite{mou_dense_2020}                   & \textbf{TMI/2020}  & 56,030,000 & 97.96 &  ---  &  ---& 98.12 &  ---  &  ---   &   ---   &   --- &  --- \\
\bottomrule
\textbf{Little U-Net}                                       &                    & 34,201     & 97.98 & 82.41 &79.81& 98.22 & 80.29 & 78.23  &  98.11  & 80.59 & 78.60  \\
\bottomrule
\textbf{Little W-Net}                                       &                    & 68,482     & \textbf{98.09} & \textbf{82.82}&\textbf{80.27}& \textbf{98.44} & \textbf{81.55} &  \textbf{79.60} & 98.24 & 81.04 & \textbf{79.11}\\
\bottomrule
\end{tabular}
\caption{Performance Comparison of methods trained/tested on DRIVE, CHASE-DB, and HRF. Best results are marked bold.}
\label{tab_results}
\end{table*}%

\subsection{Performance Evaluation}\label{vessels}
For evaluating our approach, we follow the procedure outlined in section \ref{eval_protocol}, and report AUC, DICE, and MCC values in Table \ref{tab_results}. 
For comparison purposes, we select a large set of $20$ vessel segmentation techniques published in the last years in relevant venues. We also report the performance of a standard U-Net $\phi_{3,8}$, which contains around $34,000$ parameters, and our proposed W-Net (with twice as many parameters), referred to as Little U-Net/W-Net.

As discussed above, not all techniques were trained on the same data splits for the CHASE-DB and HRF datasets. 
Our splits correspond to those used in \cite{laibacher_m2u-net_2019}, which is a model specifically designed to be efficient, and therefore contains a minimal amount of learnable parameters. 
Surprisingly, we see that the Little U-Net model already surpasses the performance of \cite{laibacher_m2u-net_2019} in all datasets, even if it has $16$ times less weights. 
The performance of the Little U-Net is overall impressive, achieving a performance on-par or superior to most of the compared techniques. 

When we analyze the performance of the Little W-Net model, we observe that it surpasses by a wide margin, both in terms of AUC and DICE score, the numbers obtained by all the other techniques. 
This is specially remarkable when considering that the Little W-Net is a far less complex model than any other approach (excluding Little U-Net). The only dataset where Little W-Net fails to reach the highest performance is HRF, which we attribute to the mismatch in training and test resolutions. The work in \cite{shin_deep_2019} , which achieves the state-of-the-art in this dataset, was trained on image patches, and it is therefore less susceptible to such mismatch. Nevertheless, the Little W-Net achieves the second best ranking in this dataset, within a short distance from \cite{shin_deep_2019}.

\subsection{Cross-dataset experiments and Domain Adapation}\label{da_eval}
From the above analysis, one could be tempted to conclude that the task of segmenting the vasculature from retinal images is relatively trivial. 
Nevertheless, the usefulness of these models remains questionable if they are not tested on data coming from sources different than the training data. 
In order to exhaustively explore this aspect of the problem, we select the W-net model trained on DRIVE images and generate predictions on up to ten different datasets (including the DRIVE test set). 
We then carry out a performance analysis similar to the one described in section \ref{eval_protocol}, and report the results in the first row of Table \ref{tab_results_da}. 
We can see how the great performance of this model on the DRIVE test set is only maintained on the STARE dataset, which is quite similar in terms of resolution and quality. However, for data arising from different distributions, this performance is rapidly degraded. 
In terms of AUC, the four worst results correspond to: 1) HRF, which has images of a much greater resolution than DRIVE, 2) LES-AV, where images are centered in the optic disc instead of in the macula, 3) AV-WIDE, which contains ultra-wield field images of markedly different aspect, and 4) UoA-DR, which has mostly pathological images of different resolutions.

\begin{figure*}[t]
\centering
\subfloat[]{\includegraphics[width = 0.49\textwidth]{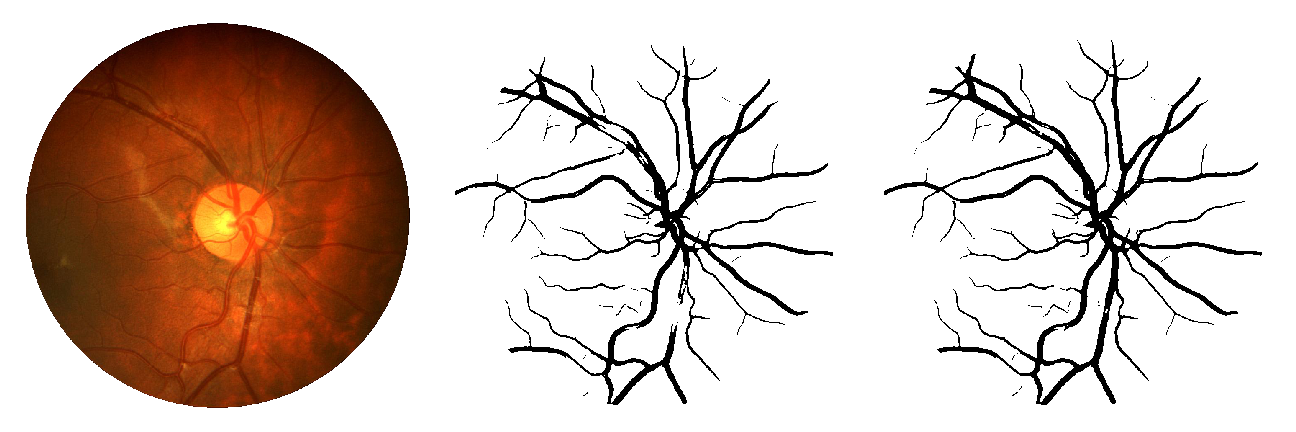}
\label{fig_deg_1}}
\hfil
\subfloat[]{\includegraphics[width = 0.49\textwidth]{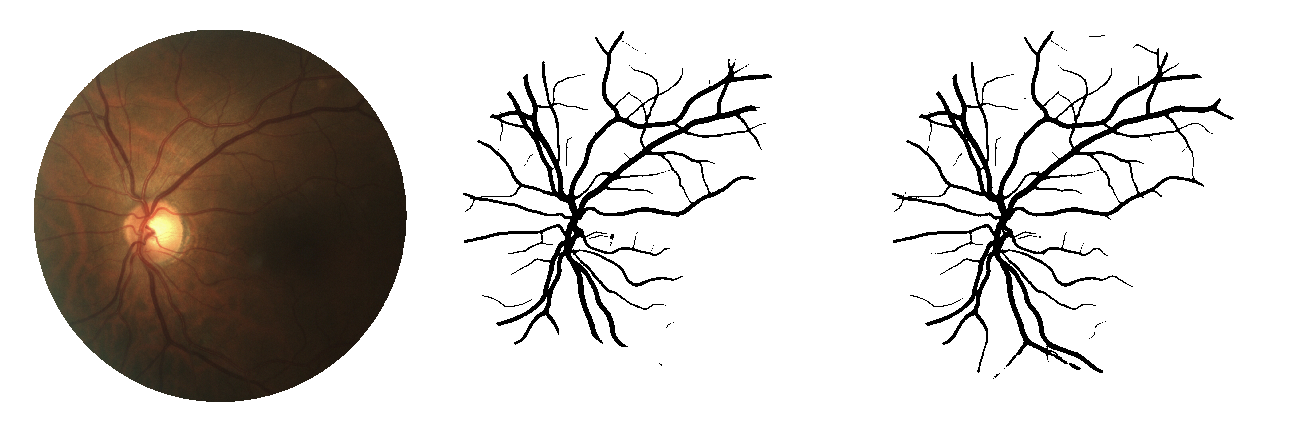}
\label{fig_deg_2}}
\hfil
\caption{The Domain Adaptation strategy from section \ref{da} recovers some missing vessels. Segmentations produced by a model trained on DRIVE (which contains macula-centered images) when using data from CHASE-DB and LES-AV (which contain OD-centered images). In (a) and (b), the retinal image (left), the segmentation by the model trained on DRIVE (center) and the one produced by the model trained on pseudo-labels (right).}
\label{fig_dr}
\end{figure*}

We then apply the strategy described in section \ref{da}: for each dataset we use the model trained on DRIVE to generate segmentations that we use as pseudo-labels to retrain the same model in an attempt to close the performance gap. 
Results of this series of experiments are displayed in the second row of Table \ref{tab_results_da}, 
where it can be seen that in almost all cases this results in an increased performance in terms of AUC, DICE score, and MCC, albeit relatively modest in some datasets. 
In any case, this implies that the retrained models have a better ability to predict vessel locations on new data.
Figure \ref{fig_dr} illustrates this for two images sampled from the CHASE-DB and the LES-AV datasets. 
Note that DRIVE does not contain optic-disc centered images. For the CHASE-DB example, we see that some broken vessels, probably due to the strong central reflex in this image, are recovered with the adapted model. In the LES-AV case, we see how an image with an uneven illumination field results in the DRIVE model missing much of the vessel pixels in the bottom area. Again, part of this vessels are successfully recovered by the adapted model.

\begin{table*}[t]  %
	\renewcommand{\arraystretch}{1}	
	\centering
\setlength\tabcolsep{2.5pt}	
\begin{tabular}{l ccc  ccc ccc ccc ccc ccc}
  & \multicolumn{3}{c}{\textbf{DRIVE}} & \multicolumn{3}{c}{\textbf{CHASE-DB}} & \multicolumn{3}{c}{\textbf{HRF}} & \multicolumn{3}{c}{\textbf{STARE}} & \multicolumn{3}{c}{\textbf{IOSTAR}}\\
\cmidrule(lr){2-4} \cmidrule(lr){5-7} \cmidrule(lr){8-10} \cmidrule(lr){11-13}  \cmidrule(lr){14-16}
\textbf{Training Set} &  AUC      &  DICE    &  MCC    &  AUC      &  DICE   &  MCC    &   AUC    &  DICE    &  MCC    &   AUC    &  DICE     &  MCC    &   AUC     &  DICE       &  MCC    \\
\midrule 
\textbf{DRIVE}        &\f{98.09}  &\f{82.82} &\f{80.27}&\s{97.22}  &\s{75.13}&\s{72.44}&\s{95.90} &\s{70.39} &\s{68.05}&\s{98.11} & \s{79.48} &\s{77.30}&  \s{97.97}& \s{78.77}   &\s{76.47} \\
\textbf{PSEUDO-L}     &\f{98.09}  &\f{82.82} &\f{80.27}&\f{97.56}  &\f{76.49}&\f{74.02}&\f{96.12} &\f{71.12} &\f{68.86}&\f{98.28} & \f{79.76} &\f{77.65}&  \f{98.06}& \f{78.95}   &\f{76.73} \\
\midrule\midrule
 & \multicolumn{3}{c}{\textbf{DRiDB}} & \multicolumn{3}{c}{\textbf{LES-AV}} & \multicolumn{3}{c}{\textbf{DR HAGIS}} & \multicolumn{3}{c}{\textbf{AV-WIDE}} & \multicolumn{3}{c}{\textbf{UoA-DR}}\\
\cmidrule(lr){2-4} \cmidrule(lr){5-7} \cmidrule(lr){8-10} \cmidrule(lr){11-13}  \cmidrule(lr){14-16}
\textbf{Training Set}  &  AUC      &  DICE    &  MCC    &  AUC      &  DICE    &  MCC    &   AUC    &  DICE      &  MCC    &   AUC     &  DICE     &  MCC    &   AUC     &  DICE      &  MCC\\
\midrule
\textbf{DRIVE}         & \s{96.17} & \f{68.45}&\f{66.62}& \s{95.45} & \s{76.60}&\s{74.32}& \s{97.17}& \s{67.92}  &\s{66.79}&\s{86.54} & \s{61.51} &\s{59.02} & \s{82.32} & \f{38.29}&\f{35.51}\\
\textbf{PSEUDO-L}      & \f{96.52} & \s{68.25}&\s{66.59}& \f{97.34} & \f{77.93}&\f{75.92}& \f{97.34}& \f{68.67}  &\f{67.49}&\f{87.64} & \f{62.46} &\f{59.97} & \f{82.71} & \s{37.68}&\s{34.97}
\end{tabular}
\caption{Our domain adaptation strategy improves results in a wide range of external test sets. First row: W-Net trained on DRIVE, second row (pseudo-l): same model fine-tuned using the strategy illustrated in Fig. \ref{da_fig}. Best metric marked bold.}
\label{tab_results_da}
\end{table*}%

\begin{table*}[t]  %
	\renewcommand{\arraystretch}{1.3}	
	\centering
\setlength\tabcolsep{12pt}	
\begin{tabular}{lc  cc cc cc}
&  & \multicolumn{2}{c}{\textbf{DRIVE}} & \multicolumn{2}{c}{\textbf{HRF}} & \multicolumn{2}{c}{\textbf{LES-AV}$^*$} \\
\cmidrule(lr){3-4} \cmidrule(lr){5-6} \cmidrule(lr){7-8} 
				&   \bfseries \# Params                  &  DICE                           &  MCC                       &  DICE             &  MCC           &   DICE  &  MCC    \\
\midrule
\cite{galdran_uncertainty-aware_2019} & $\sim$29M  &  96.31         | \textbf{96.25} &  74.79 | 25.07                   &  -----            & -----          &  \textbf{96.59}&  \textbf{70.58}  \\
\bottomrule
\cite{hemelings_arteryvein_2019}      & $\sim$5M   & \textbf{96.71} | 95.81          &  77.57 | 24.67                   &  96.88            & \textbf{76.89 }&   ---          &   ---   \\
\bottomrule
\textbf{W-Net}                        & $\sim$279K & 96.69          | 95.55          & \textbf{77.73} | \textbf{25.23}  &  \textbf{96.89}   & 76.19          &  96.46         & 70.30 \\
\bottomrule
\end{tabular}
\caption{Performance Comparison for the artery/vein segmentation task. For DRIVE, performance is reported on the entire image domain | on a ring-shaped region around the Optic Disc \cite{hemelings_arteryvein_2019}. Performance is computed using the predictions and code provided by \cite{hemelings_arteryvein_2019}. $^*$Predictions on LES-AV are generated from models trained on DRIVE.}
\label{tab_av}
\end{table*}%

\subsection{Artery/Vein Segmentation}
We also provide results for the related problem of Artery/Vein segmentation. 
It should be stressed that this is a different task than A/V classification, where the the vessel tree is assumed to be available, and the goal is to classify each vessel pixel among the two classes. 
In this case, we aim to classify each pixel in the entire image as artery, vein, or background. 
In order to account for the greater difficulty of the problem, we consider a bigger W-Net composed of two U-Nets $\phi_{4,8}$, which still contains far less weights than current A/V segmentation models \cite{galdran_uncertainty-aware_2019,hemelings_arteryvein_2019}.
In addition, we double the number of training cycles, and train with 4 classes having into account uncertain pixels, as it has been proven beneficial for this task \cite{galdran_uncertainty-aware_2019}. 

Table \ref{tab_av} shows the results of our W-Net, compared with two recent A/V segmentation techniques. 
In this section, we train our model on DRIVE and HRF, following the data splits provided in \cite{hemelings_arteryvein_2019}. 
We also show results of a cross-dataset experiment in which a model trained on DRIVE is tested on the LES-AV dataset.

\begin{figure*}[t]
\centering
\subfloat[]{\includegraphics[width = 0.43\textwidth]{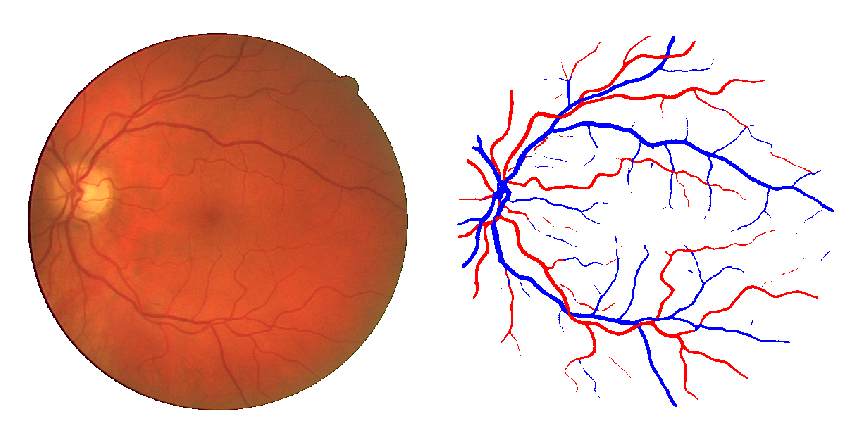}
\label{fig_deg_1}}
\hfil
\subfloat[]{\includegraphics[width = 0.43\textwidth]{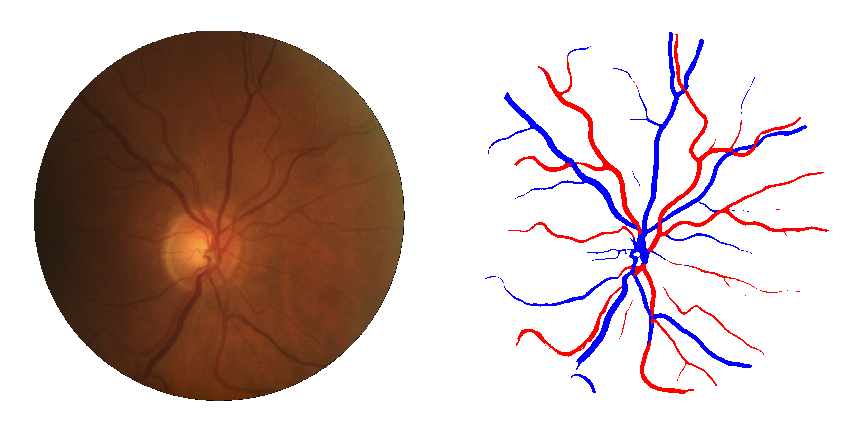}
\label{fig_deg_2}}
\hfil
\caption{Generalization ability of a W-Net trained for A/V segmentation. Results of our model trained on DRIVE and tested on (a) DRIVE, (b) LES-AV.}
\label{fig_av}
\end{figure*}

A similar trend as in Section \ref{vessels} can be observed: other models designed for the same task contain orders of magnitude more parameters than our approach, but we observe an excellent performance of the W-Net architecture: it seems competitive with the compared methods, ranking even higher than \cite{hemelings_arteryvein_2019} in terms of Dice score and higher than \cite{galdran_uncertainty-aware_2019} in terms of MCC, at a fraction of computational cost. Some qualitative results of the W-Net trained on DRIVE and tested on LES-AV are shown in Fig. \ref{fig_av}.

\begin{table*}[!b]  %
	\renewcommand{\arraystretch}{1.3}	
	\centering
\setlength\tabcolsep{12pt}	
\begin{tabular}{lc  cc cc cc}
&  & \multicolumn{2}{c}{DRIVE} & \multicolumn{2}{c}{CHASE-DB} & \multicolumn{2}{c}{HRF} \\
\cmidrule(lr){3-4} \cmidrule(lr){5-6} \cmidrule(lr){7-8} 
				&   \bfseries \# Params      &  AUC  &  DICE   &  AUC  &  DICE  &  AUC   &  DICE    \\
\midrule
\textbf{``Big'' U-Net}           & 76,213     & 98.00 & 82.53  & 98.29 & 81.09  &  98.15 &  80.73    \\
\bottomrule
\textbf{Little W-Net}            & 68,482     & \textbf{98.09} & \textbf{82.78} & \textbf{98.44} & \textbf{81.52} & \textbf{98.24} & \textbf{81.05} \\
\bottomrule
\textbf{W-Net vs U-Net}          & \textbf{-7,731} & \begin{tabular}{@{}c@{}}\textbf{+0.09} \\ p<0.05\end{tabular} & \begin{tabular}{@{}c@{}}\textbf{+0.25} \\ p<0.05\end{tabular} & \begin{tabular}{@{}c@{}}\textbf{+0.15} \\ p<0.05\end{tabular} & \begin{tabular}{@{}c@{}}\textbf{+0.43} \\ p<0.05\end{tabular} &  \begin{tabular}{@{}c@{}}\textbf{+0.09} \\ p<0.05\end{tabular}  &   \begin{tabular}{@{}c@{}}\textbf{+0.32} \\ p<0.05\end{tabular}  \\
\bottomrule
\end{tabular}
\caption{Performance comparison between a W-Net and a U-Net configured to have a comparable amount of weights. W-Net achieves higher performance, despite having slightly less parameters. Statistically significant results marked bold.}
\label{tab_ablation}
\end{table*}%

\subsection{Ablation Study: W-Net vs U-Net}
As shown in Section \ref{vessels}, the iterative structure of the W-Net architecture helps in achieving a better performance when compared to the standard U-Net. 
However, it should be noted that W-Net contains twice as many weights as the considered Little U-Net. 
Since these are two relatively small models, it might be that U-Net is simply underfitting, and all the benefits observed in Table \ref{tab_results} just come from doubling the parameters and not from any algorithmic improvement. 

In view of this, it is worth investigating the question of whether W-Net brings a significant improvement over a standard U-Net architecture. 
For this, we consider a larger U-Net $\phi_{3,12}$, which actually contains more parameters than the above W-Net ($~76K$ vs $~68K$).
To determine statistically significant differences in AUC and DICE between these two models, we train them under the exact same conditions as in Section \ref{vessels}, and after generating the corresponding predicted segmentations on each of the three test sets, we apply the bootstrap procedure as in \cite{samuelson_comparing_2006,bria_addressing_2020}. 
This is, each test set is randomly sampled with replacement 100 times so that each new set of sampled data contains the same number of examples as the original set, in the same proportion of vessel/background pixels. 
For both models, we calculate the differences in AUCs and dice scores. 
Resampling 100 times results in 100 values for performance differences. 
P-values are defined as the fraction of values that are negative or zero, corresponding to cases in which the better model in each dataset performed worse or equally than the other model. 
The statistical significance level is set to 5$\%$ and, thus, performance differences are considered statistically significant if $p < 0.05$. 
The resulting performance differences are reported in Table \ref{tab_ablation}, were we refer to the U-Net $\phi_{3,12}$ as ``Big U-Net''. 
We see that, in all cases, the larger U-Net's results are slightly better than the smaller U-Net in Table \ref{tab_results}, but the performance of the W-Net is still significantly higher, even if it has approximately $10\%$ less weights.

\subsection{Computational and Memory Requirements}
The reduced complexity of the models proposed in this paper enhance their suitability for resource-constrained scenarios, both in terms of training them and of deploying them in, \textit{e.g.}, portable devices. 
Training a little U-Net and a little W-Net to reach the performance shown in Table 2 is feasible even without a GPU. 
When training on a single GPU (GeForce RTX 2080 Ti), the training time of a little U-Net on the datasets shown in Table 2 was 24 mins (DRIVE), 22 mins (CHASE-DB) and 102 mins (HRF), whereas the little W-Net took 32 mins (DRIVE), 30 mins (CHASE-DB) and 140 mins (HRF). 
Regarding disk memory requirements, Table \ref{tab_eff} shows a comparison of both architectures with another two popular models in terms of performance vs. number of parameters/disk size. 
We see that a little U-Net, which already attains a great performance, has the lowest disk storage space (161Kb), and the top-performant W-Net takes approximately twice this space, which is still well within limits for its deployment in embedded/portable devices. 
It must be noted, however, that in both cases the inference time was slightly slower when compare to other efficient approaches, partly due to implementation of Test-Time Augmentation.

\begin{table*}[t]  %
	\renewcommand{\arraystretch}{1.3}	
	\centering
\setlength\tabcolsep{8pt}	
\begin{tabular}{lcc  cc cc cc}
&  &  & \multicolumn{2}{c}{\textbf{DRIVE}} & \multicolumn{2}{c}{\textbf{CHASEDB}} & \multicolumn{2}{c}{\textbf{HRF}} \\
\cmidrule(lr){4-5} \cmidrule(lr){6-7} \cmidrule(lr){8-9} 
				                     & \# Params & Size      & AUC       &  DICE     &  AUC     &  DICE   &  AUC   &  DICE \\
\midrule
DRIU \cite{maninis_deep_2016}        & ~15M      &    57Mb   & n/a       &  82.20    &  n/a     &  n/a    &   n/a  &   n/a \\
\bottomrule
M2U-Net \cite{laibacher_m2u-net_2019}& 0.5M      &   2.2Mb   & 97.14     &  80.91    &  97.03   &  80.06  &   n/a  &  78.14 \\
\bottomrule
Little U-Net                         & 34K       &   161Kb   & 97.98     &  82.41    &  98.22   &  80.68  &  98.11  & 80.59 \\
\bottomrule
Little W-Net                         & 68K       &   325Kb   & 98.09     &  82.82    &  98.44   &  81.55  &  98.24 &  81.04  \\
\bottomrule
\end{tabular}
\caption{Parameters and memory requirements vs performance for several retinal vessel segmentation models.}
\label{tab_eff}
\end{table*}%

\section{Conclusions}
This paper reflects on the need of constructing algorithmically complex methodologies for the task of retinal vessel segmentation. 
In a quest for squeezing an extra drop of performance on public benchmark datasets and adding certain novelty, recent works on this topic show a trend to develop overcomplicated pipelines that may not be necessary for this task. 
The first conclusion to be drawn from our work is that sometimes Occam's razor works best: minimalistic models, properly trained, can attain results that do not significantly differ from what one can achieve with more complex approaches.

Another point worth stressing is the need of rigor in evaluating retinal vessel segmentation techniques. 
Employing overly favorable train/test splits or incorrectly computing performance leads to reporting inflated metrics, which in turn saturate public benchmarks and provides a false sensation in the community that the retinal vessel segmentation problem is solved. 
Our experiments on a wide range of datasets reveal that this is not the case, and that retinal vessel segmentation is indeed an ideal area for experimenting with domain adaptation techniques. 
This is so because a) performance of models trained on a source dataset rapidly degrades when testing on a different kind of data, and b) training models to achieve high performance is cheap and fast, which enables fast experimentation of new ideas.


\bibliographystyle{unsrt} 
\bibliography{ret_vessels_refs}

\end{document}